\documentclass[twocolumn]{aastex62}
\usepackage{amsmath,amstext}
\usepackage{comment}
\usepackage{apjfonts} 
\usepackage{chngcntr}
\usepackage{float}

\newcommand{\be}{\begin{eqnarray}}
\newcommand{\ee}{\end{eqnarray}}
\newcommand{\lp}{\left(}
\newcommand{\rp}{\right)}

\begin{document}

\normalsize


\title{\Large \textbf{Shock Breakout in Dense Circumstellar Material with Application to PS1-13arp}}

\author{Annastasia Haynie}
\affiliation{Department of Physics and Astronomy, University of Southern California, Los Angeles, CA 90089, USA; ahaynie@usc.edu}
\affiliation{The Observatories of the Carnegie Institution for Science, 813 Santa Barbara St., Pasadena, CA 91101, USA}

\author{Anthony L. Piro}
\affiliation{The Observatories of the Carnegie Institution for Science, 813 Santa Barbara St., Pasadena, CA 91101, USA}

\begin{abstract}
    
    Shock breakout (SBO), the first expected electromagnetic signature of a supernova (SN), can be an important probe of the progenitors of these explosions. Unfortunately, SBO is difficult to capture with current surveys due to its brief timescale ($\lesssim 1\,$hr). However, SBO may be lengthened when dense circumstellar material (CSM) is present. Indeed, recent photometric modeling studies of SNe, as well as early spectroscopy, suggest that such dense CSM may be present more often than previously expected. If true, this should also affect the features of SBO. We present an exploration of the impact of such CSM interaction on the SBO width and luminosity using both analytic and numerical modeling, where we parameterize the CSM as a steady-state wind. We then compare this modeling to PS1-13arp, a SN that showed an early UV excess that has been argued to be SBO in dense CSM. We find PS1-13arp is well fit with a wind of mass $\sim 0.08\, M_{\odot}$ and radius $\sim 1900\, R_{\odot}$, parameters which are similar to, if not slightly less massive than, what have been inferred for Type II SNe using photometric modeling. This similarity suggests that future SBO observations of SNe~II may be easier to obtain than previously appreciated.
    
\end{abstract}


\section{Introduction}
\label{sec:introduction}
The initial electromagnetic signal of a core-collapse supernova (SN) is shock breakout (SBO). This occurs when the shock wave reaches a sufficiently low optical depth of $\tau_{\rm sbo} \approx c/v_{s}$, where $v_{s}$ is the shock speed, that radiation from the explosion can escape in front of the shock and break out of the star (\citealp{Waxman_2017}, and references therein). SBO emission can provide valuable information about the SN progenitor, such as its radius and outer density structure. However, it is difficult to observe. The radiation escapes in roughly the light crossing time, which is $\lesssim1\,$hr for a typical red supergiant \citep{Matzner_1999}. The emission also has a characteristic temperature of several $10^5$ K \citep{Falk1978, Klein1978}, which makes it dim for optically-focused, ground-based surveys.

If a massive star is embedded in dense circumstellar material (CSM), this can potentially result in a brighter and elongated SBO signal \citep{chevalier2011}, making SBO easier to study. In fact, there has been growing evidence that dense CSM may be more widely present around SNe~II progenitors than previously thought. Numerical models of SNe~II light curves performed with up-to-date red supergiant models require dense CSM to reproduce their fast rise and bright shock cooling emission (SCE) over the first $\sim 10-30\,$days \citep{unifying, measuring, moriya2017, Moriya_2018, MESA}.  These studies even argue that a lot of the variation seen in the light curves from Type IIP (plateau) to Type IIL (linear) subclasses could be due to variations in the CSM properties. High cadence early observations of SNe~II by the Transiting Exoplanet Survey Satellite (TESS) are generally brighter and show more diversity in their rise time than expected from bare red supergiant models \citep{Vallely21}, which could be due to dense CSM. Furthermore, observations of SNe~II within the first tens of hours show narrow lines again indicative of a dense wind-like environment \citep{Yaron_2017}, albeit probing more extended and less dense material than the CSM needed for light curves \citep{Dessart_2017}. X-ray emission during tens to hundreds of days after explosion also indicate dense CSM and can even help constrain its density profile \citep{Morozova18}.

The observations of SBO from these same SNe could provide an important test of these inferences of dense CSM. It could also provide complementary constraints on the mass, extent, and density profile of the CSM to better understand its currently unexplained origin. Studies of pre-explosion outbursts show that in some cases it can provide CSM with the necessary properties to explain Type II SN light curves \citep{morozova2020}. The mechanism to generate these outbursts remains unclear, although there have been many theoretical suggestions \citep{Yoon2010, David_Arnett_2011, Quataert_2012, Smith_2014, Moriya_2014, woosley2015, Fuller_2017}.

Motivated by these issues, we investigate the properties of SBO when red supergiants are surrounded by dense CSM. We highlight the role played by diffusion and the ``compactness'' of the CSM in determining the SBO properties with both analytic and numerical calculations. We then compare these results with the specific example of PS1-13arp \citep{Gezari_2015}.
An early UV excess seems to indicate that the SBO from this event was seen, however the UV peak is $\sim 50$ times longer than it should be for a normal red supergiant (RSG). Due to this, \cite{Gezari_2015} argue that this SBO requires dense CSM to be present. Using our numerical models, we can provide a better estimate of what CSM properties are needed to match the observations and how these properties relate to those previously inferred from Type II light curve studies \citep[e.g.,][]{measuring}.

In Section~\ref{sec:framework}, we first use analytic arguments to explore the impact of dense CSM on SBO signal shape. In Section~\ref{sec:numerical}, we generate a grid of numerical models that demonstrate the general dependencies that we found analytically. We compare these models to PS1-13arp to make better estimates of the CSM properties needed to understand its early UV excess. We then discuss how these properties compare to previous constraints for Type IIP and IIL SNe. In Section~\ref{sec:conclusion}, we summarize our results and discuss future work.

\section{Analytic Framework}
\label{sec:framework}

We first explore the properties of SBO in a dense CSM analytically, where it is dense in the sense that its optical depth is $\tau \gtrsim c/v_s$. In such cases, the SBO occurs within the CSM rather than the star itself. To efficiently explore a wide range of CSM properties, it is useful to parameterize the CSM as a steady-state wind \citep[e.g.,][]{Ofek_2010,chevalier2011,Moriya2011,Chugai_2007}. This reduces the CSM to two parameters, the wind radius, $R_{w}$, and mass loading factor, $D$. The mass loading factor is 
\be
    D = \frac{\dot{M}}{4\pi v_{w}},
\ee
where $\dot{M}$ is the mass loss rate in the wind and $v_{w}$ is the wind speed. The density profile of the wind is
\be
    \rho_{w} = \frac{D}{r^2}.
    \label{eq:rho_w}
\ee
In general, the dense CSM may have a more complicated density structure depending on the exact mechanism that generated it. Nevertheless, the broad properties of SBO should depend most sensitively on $R_{w}$ and the wind mass, so we save these details of different density distributions for future work. 

For this wind profile, the SBO optical depth is
\be
    \tau_{\rm sbo} = \int_{R_{\rm sbo}}^{R_{w}} \kappa \rho dr = \frac{R_{w}-R_{\rm sbo}}{R_{\rm sbo} R_{w}} \kappa D. 
    \label{eq:tau}
\ee
In their analysis of SBO in a wind, \cite{chevalier2011} identify a characteristic diffusion radius, 
\be
    R_{d} \equiv \frac{\kappa D v_{s}}{c},
\ee
which corresponds to the  radius of SBO in the limit of a wind with infinite extent. This naturally motivates us to examine two limits: $R_{w} < R_{d}$, which we refer to as the ``compact wind'' case, and $R_{w} > R_{d}$, which we call the ``extended wind'' case. We next estimate the expected luminosity and rise timescale for each of these cases.

\subsection{Compact Winds}
\label{sec:compact}
We focus mainly on the compact wind because this is the regime that best applies to the numerical models discussed in Section~\ref{sec:numerical}. For compact winds, SBO occurs near the front edge of the wind, so $R_{\rm sbo} \approx R_{w}$, and we can simplify Equation~(\ref{eq:tau}) to be 
 \be
    \tau_{\rm sbo} \approx \frac{R_{w}-R_{\rm sbo}}{R_{w}^2}\kappa D. 
    \label{eq:tau sbo}
 \ee
 We use this to find the diffusion time from the SBO depth,
 \be
    t_{d} \approx \frac{R_{w}-R_{\rm sbo}}{c}\tau_{\rm sbo} = \frac{\lp R_{w}-R_{\rm sbo} \rp^2}{R_{w}^2} \frac{\kappa D}{c}. 
    \label{eq:td}
 \ee
The shock travel time from the same depth is 
\be
    t_{s} = \frac{R_{w}-R_{\rm sbo}}{v_{s}}.
    \label{eq:ts}
\ee
SBO occurs when $t_{d} = t_{s}$, which allows us to solve for the depth of the SBO, 
\be
    R_{w} - R_{\rm sbo} = \frac{R_{w}^2 c}{\kappa D v_{s}} = \frac{R_{w}^2}{R_{d}}.
    \label{eq:sbo depth}
\ee
Substituting this condition back into Equation~(\ref{eq:tau sbo}), we find $\tau_{\rm sbo} \approx c/v_{s}$ as is typically estimated in the literature. Equation~(\ref{eq:sbo depth}) also can be rewritten as
\be
    R_{\rm sbo} = R_{w}\lp 1-\frac{R_{w}}{R_{d}}\rp, 
\ee
which confirms our earlier estimate that $R_{\rm sbo} \approx R_{w}$ for compact winds. 

The rise time of the bolometric SBO luminosity is determined by how quickly photons are able to diffuse out of the wind from the depth of SBO. Combining Equations (\ref{eq:ts}) and (\ref{eq:sbo depth}), we find 
\be
    t_{\rm rise} \approx \frac{R_{w}^2}{R_{d} v_{s}} \approx 0.1 R_{14}^2\,D_{17}^{-1}\,v_{9}^{-2}\,{\rm days}, 
    \label{eq:trise}
\ee
 where $D_{17}= D/10^{17}\,{\rm g\,cm^{-1}}$, $R_{14}= R_{w}/10^{14}\,{\rm cm}$, and $v_{9}= v_{s}/10^9\,{\rm cm\,s^{-1}}$. We assume $\kappa=0.34\,{\rm cm^2\,g^{-1}}$ for this expression, which is appropriate for electron scattering dominated material with solar composition. The energy of the SBO signal is found by integrating 
\be
    E_{\rm sbo} &=& \int_{R_{\rm sbo}}^{R_{w}} 4\pi r^2 \rho_{w} v_{s}^2 dr
    \approx \frac{4\pi c v_{s} R_{w}^2}{\kappa}\nonumber
    \\
    &\approx& 1.1\times10^{49}\, v_{9}\, R_{14}^2\,{\rm erg}.
\ee
Finally, the luminosity of the SBO signal is 
\be
    L_{\rm sbo} = \frac{E_{\rm sbo}}{t_{\rm rise}} \approx 4\pi D v_{s}^3
    \approx 1.3 \times 10^{45}\,D_{17}\,v_{9}^3\,{\rm erg\,s^{-1}}, 
    \label{eq:lsbo}
\ee
Note that the luminosity depends only on $D$ and shock speed $v_{s}$, and not directly on the wind radius $R_w$ (although the shock speed may vary slightly depending on $R_w$). 

\subsection{Extended Winds}
\label{sec:extended}
For completeness, we also summarize the analytic expectations for an extended wind, although we do not explore this case numerically. The assumption in this regime is $R_{w} \gg R_{\rm sbo}$, so that $R_{\rm sbo} \approx R_{d}$. This lets us approximate Equation~(\ref{eq:tau}) as 
\be
    \tau_{\rm sbo} \approx \frac{\kappa D}{R_{\rm sbo}}, 
\ee
and the diffusion time is, therefore, 
\be
    t_{d} \approx \frac{R_{\rm sbo}}{c} \tau_{\rm sbo}
    \approx \frac{\kappa D }{c}. 
\ee
Notice that this is independent of depth, so the diffusion time is roughly the same at any position well below $R_w$, including the SBO depth.  This is therefore the rise time of the SBO luminosity. 

The energy of the SBO is
\be
    E_{\rm sbo} &\approx& 4\pi R_{d}^2 \rho_{w}\lp R_{d} \rp v_{s}^2 R_{d}  \approx \frac{4\pi \kappa D^2 v_{s}^3}{c}\nonumber
    \\ &\approx& 1.4 \times 10^{51}\,D_{17}^2\,v_{9}^3\,{\rm erg}.
\ee
Finally, the luminosity of the shock breakout signal for an extended wind is 
\be
    L_{\rm sbo} = \frac{E_{\rm sbo}}{t_{d}} \approx 4\pi D v_{s}^3. 
\ee
Thus, the scaling for the SBO luminosity is roughly independent of whether it is extended or compact,  although $v_s$ would in detail also depend on the CSM properties.

\subsection{Light Travel Effects}
\label{sec:lt}
The calculations in Sections~\ref{sec:compact} and \ref{sec:extended} consider diffusion as the dominant rate-limiting process for determining the rise of SBO, but in certain cases we expect the light travel time, $t_{\rm lt} = R_{\rm sbo}/c$, to play an important role. Taking $t_{\rm lt} \approx R_{w}/c$ for a compact wind, and comparing this to the rise time give by Equation~(\ref{eq:trise}), we find the condition
\be
    R_{w} < \kappa D \lp\frac{v_{s}}{c}\rp^2,
    \label{eq:transition}
\ee
for when $t_{\rm lt}$ determines the rise of the SBO signal. This results in a different scaling for the luminosity, 
\be
    L_{\rm sbo} &=& \frac{E_{\rm sbo}}{t_{\rm lt}} \approx \frac{4\pi v_{s} c^2 R_{w}}{\kappa}\nonumber
    \\
    &\approx& 3.3\times 10^{45}\,v_{9}\,R_{14}\,{\rm erg\,s^{-1}}, 
\ee
which now depends on $R_{w}$ rather than $D$ as in the diffusion dominated cases before.

\subsection{Thermalization}
\label{sec:thermalization}
Under the assumption that the radiation is thermalized with the matter, we expect the temperature of the SBO signal to be roughly black body. If we adopt the conditions of a compact wind, we estimate this temperature to be 
\be
    T_{\rm BB} &\approx& \lp \frac{L_{\rm sbo}}{4\pi R_{w}^2\sigma_{\rm SB}} \rp ^{1/4}\nonumber
    \\
    &\approx& 1.2\times 10^5\,D_{17}^{1/4}\, v_{9}^{3/4}\, R_{14}^{-1/2}\,{\rm K},
    \label{eq:tbb}
\ee
where $\sigma_{\rm SB}$ is the Stefan-Boltzmann constant, so that the emission peaks in the UV. 

To check the validity of the black body assumption, we review the conditions for thermalization. The scattering opacity, $\kappa_{s}$, of the wind is generally much larger than the absorption opacity, $\kappa_{a}$. However, for thermalization to occur, sufficient absorptions must happen \citep{early}. Following the approach of \cite{1986rpa} for $\kappa_{a}\ll \kappa_{s}$, we define an effective opacity $\kappa_{\rm eff} \equiv \lp 3 \kappa_{a} \kappa_{s}\rp^{1/2}$, with an associated effective optical depth 
 \be
   \tau_{\rm eff} \approx  \kappa_{\rm eff}\rho_{w}\lp R_{\rm sbo}\rp (R_w-R_{\rm sbo}),
 \ee
 where $\tau_{\rm eff} \gtrsim 1$ is required for thermalization. To check whether this condition is satisfied at SBO, we again borrow conditions for the compact wind and let $R_{\rm sbo} \approx R_{w}$. For free-free emission, scaled to solar composition, $\kappa_{a} = 6.3\times10^{22}\,\rho T^{-7/2}\,{\rm g\,cm^{-2}}$, where $\rho$ and $T$ are in cgs units \citep{stellar1994}. Using our results from Section~\ref{sec:compact}, we find 
\be
    \tau_{\rm eff} \approx 0.24\, D_{17}^{1/16}\, v_{9}^{-37/16}\, R_{14}^{-1/8},
\ee
 where we include a factor of $7$ increase in $\rho$ when calculating $\kappa_{a}$ due to shock compression  \citep{Waxman_2017}. We therefore expect the models that we consider here to be roughly thermalized at the SBO depth, but not necessarily in the region above this. This could result in color temperature at the thermalization depth to be larger than $T_{\rm{BB}}$ by up to $\sim 20\%$ \citep[see the discussion in][]{SW2017}, Since the observed UV and optical photometric bands are on the Rayleigh-Jeans side of the spectrum, we expect a scaling of $L_\nu \propto R_w^2 \nu^2 T$, which would result in an up to $\sim 10\%$ decrease in the measured $R_{w}$. Since our main focus here is in understanding the SBO luminosity and rise time, we save a more detailed treatment of the thermalization for future work. Such a study should also explore different processes for thermalization depending on the thermodynamic conditions and wind composition.

\section{Numerical Models and PS1-13arp}
\label{sec:numerical}
To generate detailed SBO light curves that can be better compared with observations (as well as check our analytic expectations) we next consider numerical calculations. For the underlying stellar model, we use a $15\, M_{\odot}$ zero age main sequence red supergiant from the non-rotating, solar metallicity models of \texttt{KEPLER} \citep{weaver1978, woosley2007, woosley2015, sukhbold2014, sukhbold2016}. We use a single stellar model because our work focuses on the earliest stages before the underlying star is exposed and the emission is dominated by the CSM. We then stitch a wind onto the stellar model using the density profile given in Equation~(\ref{eq:rho_w}). The base of the wind sits where the wind and stellar model densities are equal. We set the density abruptly to zero at $R_w$. The temperature and composition are assumed to be constant across the CSM and taken from the point in the red supergiant model where the CSM attaches. A uniform wind speed of $10^6\, {\rm cm\, s^{-1}}$ is added to the CSM to prevent any fallback of material, but this exact choice does not make any noticeable differences because the velocity of the material is reset when the SN shock passes through.

The models are then exploded with the open-source code \texttt{SNEC} \citep{SNEC}. The inner $1.4\, M_{\odot}$ is excised from the model to form a neutron star and we use a ``thermal bomb'' mechanism with an energy injection of $10^{51}\,{\rm erg}$. The compositional profiles are smoothed with a ``boxcar'' approach and opacity floors are adopted, in both cases with the same parameters as in \cite{SNEC}.

Due to the large radius of the red supergiant plus CSM, light from different points on the surface of the star must travel slightly different distances to the observer. To fully incorporate this time delay, which effectively smears out the luminosity peak, we calculate the local light travel time at each point on the the star,
\be
    t_{\rm lt} = t - R_{\rm ph}(t_{\rm lt})(1- \cos{\theta})/c,
    \label{eq:tlt}
\ee    
where $t$ is the current time in the simulation, $R_{\rm ph}$ is the photospheric radius, defined as where the optical depth $\tau=2/3$ in the simulation, and $\theta$ is the latitude measured from the equator. Note that $t_{\rm lt}$ appears on both sides of Equation~(\ref{eq:tlt}), so for a given $t$ and $\theta$ we solve for $t_{\rm lt}$ iteratively. We then integrate over the surface of the star
\be
    L(t) = \int_{0}^{\pi/2} L(t_{\rm lb})\sin{\theta}d\theta,
\ee
to get the observed luminosity at any given time.

\begin{figure}
\includegraphics[width=0.45\textwidth]{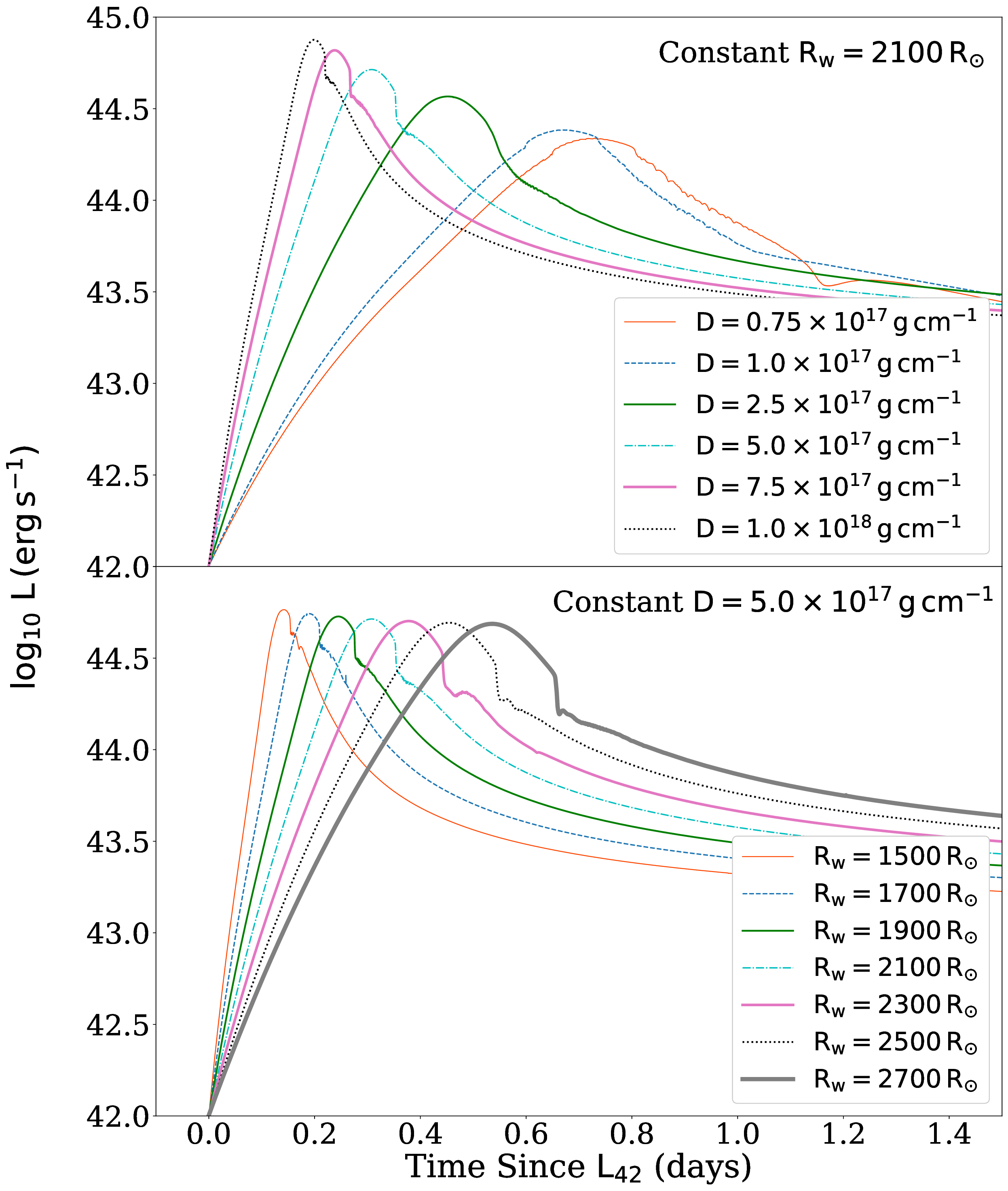}
\caption{(Top) Numerical bolometric light curves of models with fixed $R_{w} = 2100\,R_{\odot}$ and varying $D$. These calculations follow our analytic estimates, namely $t_{\rm rise} \propto 1/D$, Equation~(\ref{eq:trise}), and $L \propto D$, Equation~(\ref{eq:lsbo}), so as $D$ increases, models are narrower and brighter. (Bottom) Numerical bolometric light curves of models with fixed $D = 5.0 \times 10^{17}\, {\rm g}\,{\rm cm}^{-1}$ and varying $R_{w}$. Again from our analytic exploration, we expect $t_{\rm rise} \propto R_{w}^2$, making models wider as $R_{w}$ increases, while $L_{\rm sbo}$ does not depend strongly on $R_{w}$. Time on the x-axis is the time since the light curve reached a luminosity of $L_{42} = 10^{42}\, \mathbf{\mathrm{erg\, s^{-1}}}$.}
\label{fig:fig1}
\end{figure}

Figure~\ref{fig:fig1} compares the bolometric luminosity of a selection of models for varying values of $R_{w}$ and $D$ with a fixed explosion energy of $10^{51}\,{\rm erg}$. The top panel shows models of constant $R_{w} = 2100\,R_\odot$ with a range of $D$ values. The light curves are narrower and brighter for models with increasing $D$. This roughly follows the dependence we expect from Equations~(\ref{eq:trise}) and (\ref{eq:lsbo}), although the numerical luminosity has a somewhat weaker dependence due to the slight inverse relation between shock velocity and $D$. The shock will naturally be slower in a more dense wind, but not by enough to impact the overall positive correlation we find between luminosity and $D$. The bottom panel shows that for fixed $D = 5.0\times 10^{17}\, {\rm g}\, {\rm cm}^{-1}$, light curves get wider as $R_w$ increases, but the luminosity stays largely unchanged.  These dependencies again roughly follow the analytic scalings found in Section~\ref{sec:compact}. We do not perform a more quantitative comparison between the analytic scalings and numerical calculations since in detail the shock velocity and opacity will have variations across the CSM, issues that we will explore in more detail in future work.


\begin{figure}
\includegraphics[width=0.46\textwidth]{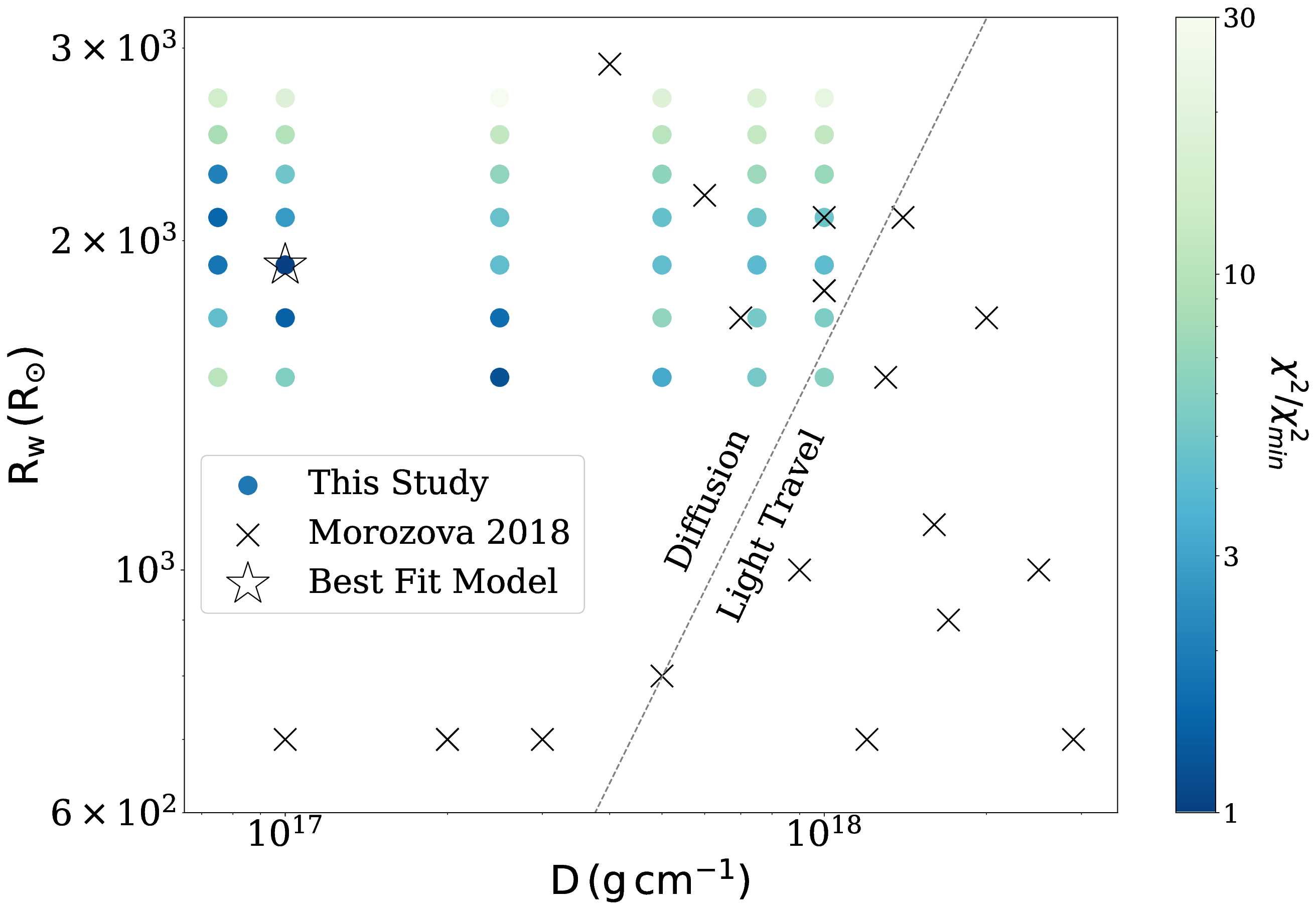}
\caption{Filled circled represent the $D$ and $R_w$ values of each of the models in our grid. The color of each point indicates the goodness of fit to PS1-13arp as the ratio of the $\chi^2$ value of that model to the $\chi_{\rm min}^2 = 17.7$. The best fit model is indicated by the star outline. For each point, we use the best fitting energy value (which corresponds to either $6 \times 10^{50}$ or $10^{51}\,$erg).  Black crosses represent the CSM parameters inferred for 20 SNe~II fit in \cite{measuring}. The gray dashed line marks the transition between where the rise time is dominated by either diffusion or light travel effects as given by Equation~(\ref{eq:transition}).}
\label{fig:fig2}
\end{figure}

We next compare our SBO models to the early emission of PS1-13arp to better understand what CSM properties are needed to replicate this event. To do this, we build a grid of 168 models in which we vary $R_w$, $D$, and the explosion energy, but keep the same underlying stellar model as above. The $R_w$ and $D$ values we consider are represented by the filled circles in Figure~\ref{fig:fig2}. For each of these models, we consider explosion energies of $3\times10^{50}$, $6\times10^{50}$, $10^{51}$, and $3\times10^{51}\,{\rm erg}$. Each model is exploded and tracked out to $\sim 23$ days to fully capture the SBO decline and transition into SCE. This length of time ensures that we have sufficient data points in each of the {\it g}-, {\it r}-, {\it i}-, and {\it z}-bands to appropriately fit the PS1-13arp data. This also helps to mitigate any biases in our fitting due to differences between the color and black body temperatures as discussed in ~\ref{sec:thermalization}. This is because at these later times in the SCE phase thermalization is expected to be more complete \citep{measuring}. We also scale the time and temperature of our models with a redshift of $z = 0.1665$, as is inferred for this event \citep{Gezari_2015}.

The goodness of fit for each model is determined by the $\chi^2$ value it yields when compared to the data with each of the 21 data points weighted equally. In principle, we could have just fit the SBO over the first $\sim5\,{\rm days}$, but including the SCE emission over the next $\sim20\,{\rm days}$ helps us double check our inferences about the presence of dense CSM and also better constrain the CSM parameters. The key is to not extend the fit so long as to have the luminosity start depending on the underlying progenitor, which we were careful not to do. Since the time of shock breakout for PS1-13arp could have occurred anywhere between the last non-detection and first detection in the NUV band, we also fit for the time of explosion. When fitting the explosion energy, we find that for any combination of $R_{w}$ and $D$, the best fit energy is always either $6 \times 10^{50}$ or $10^{51}\,{\rm erg}$. The minimum $\chi^2$ across the possible explosion times and energies is the reported $\chi^2$ for that model, which is the value utilized to compare models to each other.

The color of each filled circle in Figure~\ref{fig:fig2} represents the $\chi^2$ fits of each model with respect to the overall $\chi_{\rm min}^2 = 17.7$. The best fit model has parameters of $R_w = 1900\, R_{\odot}$ and $D = 10^{17}\,{\rm g\,cm^{-1}}$, which corresponds to a CSM mass of $\approx 0.08\, M_{\odot}$. The colors show a small degeneracy between $R_{w}$ and $D$, as has been noted in previous work \citep{unifying, measuring}. This is because the mass of the CSM is given by
\be
    M_w = 4\pi D(R_w-R_*),
\ee
where $R_*$ is the radius of the red supergiant, and the shape of the light curve is most sensitive to this mass.

\begin{figure}
\includegraphics[width=0.46\textwidth]{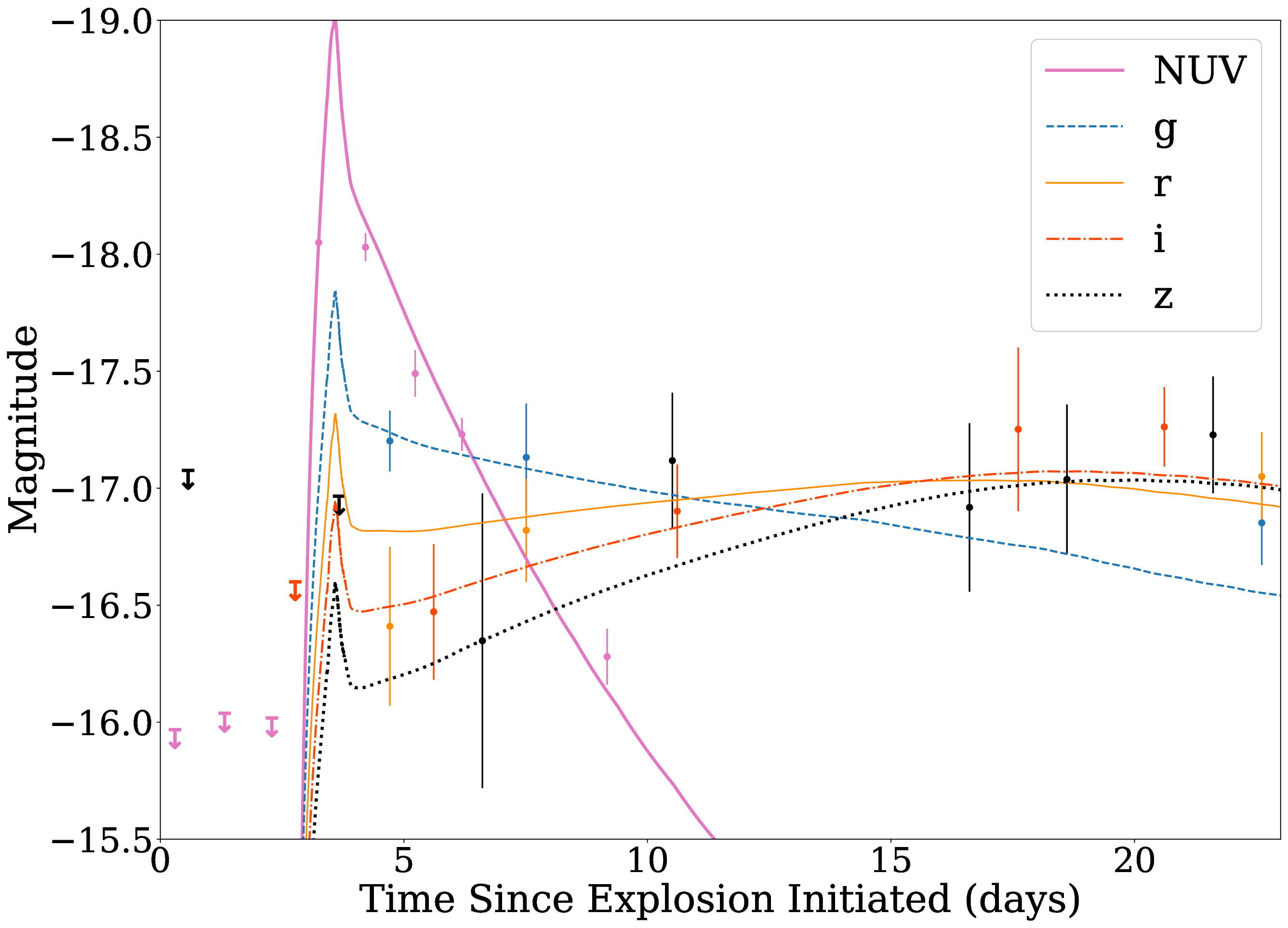}
\caption{Best fit model (solid lines) to the photometric data (points with error bars) of PS1-13arp in GALEX NUV (pink, solid), and PS1 bands \textit{g} (blue, dashed), \textit{r} (orange , solid), \textit{i} (red , dot dashed), and \textit{z} ( black, dotted). Downward arrows indicate upper limits , which are are not included in the fitting process (although they constrain the explosion time).  Time on the x-axis is the time since the explosion was initiated in SNEC. Parameters of this model are $R_{w} = 1900\, R_{\odot}$ and $D = 10^{17}\, {\rm g}\ {\rm cm}^{-1}$, which results in a total wind mass of $\approx0.08\,M_\odot$}
\label{fig:fig3}
\end{figure}

Figure~\ref{fig:fig3} shows the photometric light curves of the best fit model overlaid with PS1-13arp data. Both $R_w$ and $D$ for this model are roughly within a factor of $\sim 2$ of the estimates made in \cite{Gezari_2015}. However, \cite{Gezari_2015} use the extended wind regime to estimate their parameters, while our numerical fits demonstrate that PS1-13arp is well within the compact wind regime. Given our analytic estimates outlined in Section~\ref{sec:framework}, we rule out the possibility of the PS1-13arp data also being well fit by an extended wind for two reasons. First, attempting to match the rise time in the extended regime results in a peak bolometric luminosity nearly an order of magnitude too dim. Likewise, matching the peak luminosity would cause the light curve to rise too slowly.


Given that the bright early emission from PS1-13arp argues for this dense CSM, it is natural to ask how exceptional this event is in comparison to other SNe~II. Unfortunately, the vast majority of these events do not have the same early UV coverage of PS1-13arp. CSM properties can be inferred from other observational signatures, such as the SCE over the first $\sim30\,{\rm days}$ \citep{unifying}. To compare PS1-13arp to this other method for measuring the CSM, in Figure~\ref{fig:fig2} we overlay the CSM properties inferred for 20 SNe~II by \cite{measuring} as black crosses. The CSM in these events tend to have smaller $R_w$ and higher $D$ than PS1-13arp, but are not so inconsistent that they seem to be fundamentally different kinds of events. If the SBO of these events would have been observed, the scalings for $L_{\rm sbo}$ and $t_{\rm rise}$ imply they would have been brighter but with faster rise times. This may make SBO still challenging to observe unless a survey has a sufficiently high cadence.

\section{Discussion and Conclusion}
\label{sec:conclusion}

In this work, we investigated the impact of dense CSM on the SBO signal of core-collapse SNe. We generally find that the CSM makes SBO brighter and longer lasting than when it is just from a bare RSG, and both of these factors can make SBO potentially easier to observe.  Our analytic work highlights the important differences between the compact and extended regimes (as previously discussed by \citealp{chevalier2011}), with the compact regime appearing most relevant for observed events. Equations~(\ref{eq:trise}) and (\ref{eq:lsbo}) provide analytic estimates for how the rise time and luminosity depend on the CSM properties in this case, respectively. Somewhat paradoxically, in this compact case the rise time varies inversely with the wind density. This is because as the density increases, the SBO occurs closer to the edge of the wind. In either case, the SBO luminosity is roughly independent of the wind radius. These general scalings were then confirmed and explored in further detail using numerical explosion calculations (see Figure~\ref{fig:fig1}). 

We also compared our models with PS1-13arp, which has a bright, elongated SBO signal, previously discussed as potentially being due to SBO within dense CSM. Using our numerical models, we confirm that this is a plausible explanation for PS1-13arp's early, extended UV excess. The event is best fit by a wind with $R_{w} = 1900\, R_{\odot}$ and $D= 10^{17}\, {\rm g}\ {\rm cm}^{-1}$, as shown in Figures~\ref{fig:fig2} and \ref{fig:fig3}.   Our fit shows that the SBO of PS1-13arp is in the compact regime, which was not accounted for in previous studies of this event. We rule out the possibility of an extended wind, which cannot reproduce both the fast rise and bright peak of the data. Also in Figure~\ref{fig:fig2}, we compare the parameters we fit for PS1-13arp with SNe~II that have CSM parameters inferred from SCE \citep[black crosses,][]{measuring}. PS1-13arp's CSM appears to have a somewhat larger radius and smaller density in comparison to this larger sample, but overall the properties are not too dissimilar.  This may suggest that the mechanism for producing the CSM in all these events may be related, although more work is needed to compare the range of CSM properties expected from any given scenario \citep[e.g.,][]{Smith_2014,Fuller_2017} with the range of CSM properties inferred from observations.

Besides our results for the specific case of PS1-13arp, our work has broader implications for future studies of SBO from RSGs. SBO has long been a sought after signal to measure from core-collapse SNe because of the potentially unique information it can provide about the progenitors. Unfortunately, as mentioned earlier, this has been difficult historically due to the short timescale and high temperatures of SBO. If CSM is as ubiquitous as recent studies suggest, the future outlook becomes more optimistic. In particular, future space-based UV surveys may find significantly more SBO into CSM than might be expected from typical SBO calculations from bare RSGs \citep[e.g.,][]{cutie, guci}. The typical cadence times of $\sim90\,$min would be well-suited for resolving the SBO light curve. Of course such measurements would no longer probe the surface properties of RSG progenitors, but instead probe the CSM itself.  The Rubin Observatory \citep{LSST} will eventually provide a history of activity for nearby RSGs over the years before they explode as SNe. Combining such observations with measurements of SBO through CSM will allow us develop a more complete picture of the final weeks to months of the lives of massive stars. In this way, we can finally understand why mass loss apparently becomes so violent during the last stages of stellar evolution.

Further theoretical developments are also needed to fully interpret such observations. Although our work is best for characterizing the bolometric luminosity and rise time, incomplete thermalization means that our current black body approximation underestimates the observed radiation field (as discussed in Section~\ref{sec:thermalization}). A more detailed treatment of thermalization, and its dependence on the density profile and composition, are therefore needed.  Also, referring back to Figure~\ref{fig:fig2}, we see that many of the events studied by \cite{measuring} fall around the boundary between the diffusion and light travel dominated cases. It is thus important to better  understand what physical properties of the CSM can be best constrained if many observed events fall into this regime.

\acknowledgments
We thank Viktoriya Morozova for sharing an initial grid of RSG + CSM models, and Suvi Gezari for lending the PS1-13arp data and her assistance with interpreting the observations. We also thank Roger Chevalier and Takashi Moriya for helpful feedback on a previous draft of this manuscript. A.H. acknowledges support from the USC-Carnegie Fellowship. A.L.P. acknowledges financial support from a Scialog award made by the Research Corporation for Science Advancement.

\bibliographystyle{yahapj}

\begin{thebibliography}{}
\providecommand\natexlab[1]{#1}
\providecommand\JournalTitle[1]{#1}

\bibitem[{{Arnett} \& {Meakin}(2011)}]{David_Arnett_2011}
{Arnett}, W.~D., \& {Meakin}, C. 2011,
  \href{http://dx.doi.org/10.1088/0004-637X/733/2/78}{\JournalTitle{\apj}, 733,
  78}

\bibitem[{{Cenko}(2019)}]{guci}
{Cenko}, S.~B. 2019, in American Astronomical Society Meeting Abstracts, Vol.
  234, American Astronomical Society Meeting Abstracts \#234, 212.03

\bibitem[{{Cenko} {et~al.}(2017){Cenko}, {Bellm}, {Gal-Yam}, {Gezari},
  {Gorjian}, {Jewell}, {Kruk}, {Kulkarni}, {Mushotzky}, {Nikzad}, {Piro},
  {Waxman}, \& {Ofek}}]{cutie}
{Cenko}, S.~B., {Bellm}, E.~C., {Gal-Yam}, A., {et~al.} 2017, in American
  Astronomical Society Meeting Abstracts, Vol. 229, American Astronomical
  Society Meeting Abstracts \#229, 328.04

\bibitem[{{Chevalier} \& {Irwin}(2011)}]{chevalier2011}
{Chevalier}, R.~A., \& {Irwin}, C.~M. 2011,
  \href{http://dx.doi.org/10.1088/2041-8205/729/1/L6}{\JournalTitle{\apjl},
  729, L6}

\bibitem[{{Chugai} {et~al.}(2007){Chugai}, {Chevalier}, \&
  {Utrobin}}]{Chugai_2007}
{Chugai}, N.~N., {Chevalier}, R.~A., \& {Utrobin}, V.~P. 2007,
  \href{http://dx.doi.org/10.1086/518160}{\JournalTitle{\apj}, 662, 1136}

\bibitem[{{Dessart} {et~al.}(2017){Dessart}, {John Hillier}, {Yoon}, {Waldman},
  \& {Livne}}]{Dessart_2017}
{Dessart}, L., {John Hillier}, D., {Yoon}, S.-C., {Waldman}, R., \& {Livne}, E.
  2017,
  \href{http://dx.doi.org/10.1051/0004-6361/201730873}{\JournalTitle{\aap},
  603, A51}

\bibitem[{{Falk}(1978)}]{Falk1978}
{Falk}, S.~W. 1978,
  \href{http://dx.doi.org/10.1086/182810}{\JournalTitle{\apjl}, 225, L133}

\bibitem[{{Fuller}(2017)}]{Fuller_2017}
{Fuller}, J. 2017,
  \href{http://dx.doi.org/10.1093/mnras/stx1314}{\JournalTitle{\mnras}, 470,
  1642}

\bibitem[{{Gezari} {et~al.}(2015){Gezari}, {Jones}, {Sanders}, {Soderberg},
  {Hung}, {Heinis}, {Smartt}, {Rest}, {Scolnic}, {Chornock}, {Berger}, {Foley},
  {Huber}, {Price}, {Stubbs}, {Riess}, {Kirshner}, {Smith}, {Wood-Vasey},
  {Schiminovich}, {Martin}, {Burgett}, {Chambers}, {Flewelling}, {Kaiser},
  {Tonry}, \& {Wainscoat}}]{Gezari_2015}
{Gezari}, S., {Jones}, D.~O., {Sanders}, N.~E., {et~al.} 2015,
  \href{http://dx.doi.org/10.1088/0004-637X/804/1/28}{\JournalTitle{\apj}, 804,
  28}

\bibitem[{{Hansen} \& {Kawaler}(1994)}]{stellar1994}
{Hansen}, C.~J., \& {Kawaler}, S.~D. 1994, {Stellar Interiors. Physical
  Principles, Structure, and Evolution.}

\bibitem[{{Klein} \& {Chevalier}(1978)}]{Klein1978}
{Klein}, R.~I., \& {Chevalier}, R.~A. 1978,
  \href{http://dx.doi.org/10.1086/182740}{\JournalTitle{\apjl}, 223, L109}

\bibitem[{{LSST Science Collaboration} {et~al.}(2009){LSST Science
  Collaboration}, {Abell}, {Allison}, {Anderson}, {Andrew}, {Angel}, {Armus},
  {Arnett}, {Asztalos}, {Axelrod}, {Bailey}, {Ballantyne}, {Bankert},
  {Barkhouse}, {Barr}, {Barrientos}, {Barth}, {Bartlett}, {Becker}, {Becla},
  {Beers}, {Bernstein}, {Biswas}, {Blanton}, {Bloom}, {Bochanski}, {Boeshaar},
  {Borne}, {Bradac}, {Brandt}, {Bridge}, {Brown}, {Brunner}, {Bullock},
  {Burgasser}, {Burge}, {Burke}, {Cargile}, {Chandrasekharan}, {Chartas},
  {Chesley}, {Chu}, {Cinabro}, {Claire}, {Claver}, {Clowe}, {Connolly}, {Cook},
  {Cooke}, {Cooray}, {Covey}, {Culliton}, {de Jong}, {de Vries}, {Debattista},
  {Delgado}, {Dell'Antonio}, {Dhital}, {Di Stefano}, {Dickinson}, {Dilday},
  {Djorgovski}, {Dobler}, {Donalek}, {Dubois-Felsmann}, {Durech},
  {Eliasdottir}, {Eracleous}, {Eyer}, {Falco}, {Fan}, {Fassnacht}, {Ferguson},
  {Fernandez}, {Fields}, {Finkbeiner}, {Figueroa}, {Fox}, {Francke}, {Frank},
  {Frieman}, {Fromenteau}, {Furqan}, {Galaz}, {Gal-Yam}, {Garnavich},
  {Gawiser}, {Geary}, {Gee}, {Gibson}, {Gilmore}, {Grace}, {Green}, {Gressler},
  {Grillmair}, {Habib}, {Haggerty}, {Hamuy}, {Harris}, {Hawley}, {Heavens},
  {Hebb}, {Henry}, {Hileman}, {Hilton}, {Hoadley}, {Holberg}, {Holman},
  {Howell}, {Infante}, {Ivezic}, {Jacoby}, {Jain}, {R}, {Jedicke}, {Jee},
  {Garrett Jernigan}, {Jha}, {Johnston}, {Jones}, {Juric}, {Kaasalainen},
  {Styliani}, {Kafka}, {Kahn}, {Kaib}, {Kalirai}, {Kantor}, {Kasliwal},
  {Keeton}, {Kessler}, {Knezevic}, {Kowalski}, {Krabbendam}, {Krughoff},
  {Kulkarni}, {Kuhlman}, {Lacy}, {Lepine}, {Liang}, {Lien}, {Lira}, {Long},
  {Lorenz}, {Lotz}, {Lupton}, {Lutz}, {Macri}, {Mahabal}, {Mandelbaum},
  {Marshall}, {May}, {McGehee}, {Meadows}, {Meert}, {Milani}, {Miller},
  {Miller}, {Mills}, {Minniti}, {Monet}, {Mukadam}, {Nakar}, {Neill}, {Newman},
  {Nikolaev}, {Nordby}, {O'Connor}, {Oguri}, {Oliver}, {Olivier}, {Olsen},
  {Olsen}, {Olszewski}, {Oluseyi}, {Padilla}, {Parker}, {Pepper}, {Peterson},
  {Petry}, {Pinto}, {Pizagno}, {Popescu}, {Prsa}, {Radcka}, {Raddick},
  {Rasmussen}, {Rau}, {Rho}, {Rhoads}, {Richards}, {Ridgway}, {Robertson},
  {Roskar}, {Saha}, {Sarajedini}, {Scannapieco}, {Schalk}, {Schindler},
  {Schmidt}, {Schmidt}, {Schneider}, {Schumacher}, {Scranton}, {Sebag},
  {Seppala}, {Shemmer}, {Simon}, {Sivertz}, {Smith}, {Allyn Smith}, {Smith},
  {Spitz}, {Stanford}, {Stassun}, {Strader}, {Strauss}, {Stubbs}, {Sweeney},
  {Szalay}, {Szkody}, {Takada}, {Thorman}, {Trilling}, {Trimble}, {Tyson}, {Van
  Berg}, {Vanden Berk}, {VanderPlas}, {Verde}, {Vrsnak}, {Walkowicz},
  {Wandelt}, {Wang}, {Wang}, {Warner}, {Wechsler}, {West}, {Wiecha},
  {Williams}, {Willman}, {Wittman}, {Wolff}, {Wood-Vasey}, {Wozniak}, {Young},
  {Zentner}, \& {Zhan}}]{LSST}
{LSST Science Collaboration}, {Abell}, P.~A., {Allison}, J., {et~al.} 2009,
  \JournalTitle{arXiv e-prints}, arXiv:0912.0201

\bibitem[{{Matzner} \& {McKee}(1999)}]{Matzner_1999}
{Matzner}, C.~D., \& {McKee}, C.~F. 1999,
  \href{http://dx.doi.org/10.1086/306571}{\JournalTitle{\apj}, 510, 379}

\bibitem[{{Moriya} {et~al.}(2011){Moriya}, {Tominaga}, {Blinnikov}, {Baklanov},
  \& {Sorokina}}]{Moriya2011}
{Moriya}, T., {Tominaga}, N., {Blinnikov}, S.~I., {Baklanov}, P.~V., \&
  {Sorokina}, E.~I. 2011,
  \href{http://dx.doi.org/10.1111/j.1365-2966.2011.18689.x}{\JournalTitle{\mnras},
  415, 199}

\bibitem[{{Moriya} \& {Langer}(2015)}]{Moriya_2014}
{Moriya}, T.~J., \& {Langer}, N. 2015,
  \href{http://dx.doi.org/10.1051/0004-6361/201424957}{\JournalTitle{\aap},
  573, A18}

\bibitem[{{Moriya} {et~al.}(2018){Moriya}, {Sorokina}, \&
  {Chevalier}}]{Moriya_2018}
{Moriya}, T.~J., {Sorokina}, E.~I., \& {Chevalier}, R.~A. 2018,
  \href{http://dx.doi.org/10.1007/s11214-018-0493-6}{\JournalTitle{\ssr}, 214,
  59}

\bibitem[{{Moriya} {et~al.}(2017){Moriya}, {Yoon}, {Gr{\"a}fener}, \&
  {Blinnikov}}]{moriya2017}
{Moriya}, T.~J., {Yoon}, S.-C., {Gr{\"a}fener}, G., \& {Blinnikov}, S.~I. 2017,
  \href{http://dx.doi.org/10.1093/mnrasl/slx056}{\JournalTitle{\mnras}, 469,
  L108}

\bibitem[{{Morozova} {et~al.}(2015){Morozova}, {Ott}, \& {Piro}}]{SNEC}
{Morozova}, V., {Ott}, C.~D., \& {Piro}, A.~L. 2015, {SNEC: SuperNova Explosion
  Code}

\bibitem[{{Morozova} {et~al.}(2020){Morozova}, {Piro}, {Fuller}, \& {Van
  Dyk}}]{morozova2020}
{Morozova}, V., {Piro}, A.~L., {Fuller}, J., \& {Van Dyk}, S.~D. 2020,
  \href{http://dx.doi.org/10.3847/2041-8213/ab77c8}{\JournalTitle{\apjl}, 891,
  L32}

\bibitem[{{Morozova} {et~al.}(2017){Morozova}, {Piro}, \& {Valenti}}]{unifying}
{Morozova}, V., {Piro}, A.~L., \& {Valenti}, S. 2017,
  \href{http://dx.doi.org/10.3847/1538-4357/aa6251}{\JournalTitle{\apj}, 838,
  28}

\bibitem[{{Morozova} {et~al.}(2018){Morozova}, {Piro}, \&
  {Valenti}}]{measuring}
---. 2018,
  \href{http://dx.doi.org/10.3847/1538-4357/aab9a6}{\JournalTitle{\apj}, 858,
  15}

\bibitem[{{Morozova} \& {Stone}(2018)}]{Morozova18}
{Morozova}, V., \& {Stone}, J.~M. 2018,
  \href{http://dx.doi.org/10.3847/1538-4357/aae2b3}{\JournalTitle{\apj}, 867,
  4}

\bibitem[{{Nakar} \& {Sari}(2010)}]{early}
{Nakar}, E., \& {Sari}, R. 2010,
  \href{http://dx.doi.org/10.1088/0004-637X/725/1/904}{\JournalTitle{\apj},
  725, 904}

\bibitem[{{Ofek} {et~al.}(2010){Ofek}, {Rabinak}, {Neill}, {Arcavi}, {Cenko},
  {Waxman}, {Kulkarni}, {Gal-Yam}, {Nugent}, {Bildsten}, {Bloom}, {Filippenko},
  {Forster}, {Howell}, {Jacobsen}, {Kasliwal}, {Law}, {Martin}, {Poznanski},
  {Quimby}, {Shen}, {Sullivan}, {Dekany}, {Rahmer}, {Hale}, {Smith},
  {Zolkower}, {Velur}, {Walters}, {Henning}, {Bui}, \& {McKenna}}]{Ofek_2010}
{Ofek}, E.~O., {Rabinak}, I., {Neill}, J.~D., {et~al.} 2010,
  \href{http://dx.doi.org/10.1088/0004-637X/724/2/1396}{\JournalTitle{\apj},
  724, 1396}

\bibitem[{{Paxton} {et~al.}(2018){Paxton}, {Schwab}, {Bauer}, {Bildsten},
  {Blinnikov}, {Duffell}, {Farmer}, {Goldberg}, {Marchant}, {Sorokina},
  {Thoul}, {Townsend}, \& {Timmes}}]{MESA}
{Paxton}, B., {Schwab}, J., {Bauer}, E.~B., {et~al.} 2018,
  \href{http://dx.doi.org/10.3847/1538-4365/aaa5a8}{\JournalTitle{\apjs}, 234,
  34}

\bibitem[{{Quataert} \& {Shiode}(2012)}]{Quataert_2012}
{Quataert}, E., \& {Shiode}, J. 2012,
  \href{http://dx.doi.org/10.1111/j.1745-3933.2012.01264.x}{\JournalTitle{\mnras},
  423, L92}

\bibitem[{{Rybicki} \& {Lightman}(1986)}]{1986rpa}
{Rybicki}, G.~B., \& {Lightman}, A.~P. 1986, {Radiative Processes in
  Astrophysics}

\bibitem[{{Sapir} \& {Waxman}(2017)}]{SW2017}
{Sapir}, N., \& {Waxman}, E. 2017,
  \href{http://dx.doi.org/10.3847/1538-4357/aa64df}{\JournalTitle{\apj}, 838,
  130}

\bibitem[{{Smith} \& {Arnett}(2014)}]{Smith_2014}
{Smith}, N., \& {Arnett}, W.~D. 2014,
  \href{http://dx.doi.org/10.1088/0004-637X/785/2/82}{\JournalTitle{\apj}, 785,
  82}

\bibitem[{{Sukhbold} {et~al.}(2016){Sukhbold}, {Ertl}, {Woosley}, {Brown}, \&
  {Janka}}]{sukhbold2016}
{Sukhbold}, T., {Ertl}, T., {Woosley}, S.~E., {Brown}, J.~M., \& {Janka}, H.~T.
  2016,
  \href{http://dx.doi.org/10.3847/0004-637X/821/1/38}{\JournalTitle{\apj}, 821,
  38}

\bibitem[{{Sukhbold} \& {Woosley}(2014)}]{sukhbold2014}
{Sukhbold}, T., \& {Woosley}, S.~E. 2014,
  \href{http://dx.doi.org/10.1088/0004-637X/783/1/10}{\JournalTitle{\apj}, 783,
  10}

\bibitem[{{Vallely} {et~al.}(2021){Vallely}, {Kochanek}, {Stanek}, {Fausnaugh},
  \& {Shappee}}]{Vallely21}
{Vallely}, P.~J., {Kochanek}, C.~S., {Stanek}, K.~Z., {Fausnaugh}, M., \&
  {Shappee}, B.~J. 2021,
  \href{http://dx.doi.org/10.1093/mnras/staa3675}{\JournalTitle{\mnras}, 500,
  5639}

\bibitem[{{Waxman} \& {Katz}(2017)}]{Waxman_2017}
{Waxman}, E., \& {Katz}, B. 2017, {Shock Breakout Theory}, ed. A.~W. {Alsabti}
  \& P.~{Murdin}, 967

\bibitem[{{Weaver} {et~al.}(1978){Weaver}, {Zimmerman}, \&
  {Woosley}}]{weaver1978}
{Weaver}, T.~A., {Zimmerman}, G.~B., \& {Woosley}, S.~E. 1978,
  \href{http://dx.doi.org/10.1086/156569}{\JournalTitle{\apj}, 225, 1021}

\bibitem[{{Woosley} \& {Heger}(2007)}]{woosley2007}
{Woosley}, S.~E., \& {Heger}, A. 2007,
  \href{http://dx.doi.org/10.1016/j.physrep.2007.02.009}{\JournalTitle{\physrep},
  442, 269}

\bibitem[{{Woosley} \& {Heger}(2015)}]{woosley2015}
---. 2015,
  \href{http://dx.doi.org/10.1088/0004-637X/810/1/34}{\JournalTitle{\apj}, 810,
  34}

\bibitem[{{Yaron} {et~al.}(2017){Yaron}, {Perley}, {Gal-Yam}, {Groh}, {Horesh},
  {Ofek}, {Kulkarni}, {Sollerman}, {Fransson}, {Rubin}, {Szabo}, {Sapir},
  {Taddia}, {Cenko}, {Valenti}, {Arcavi}, {Howell}, {Kasliwal}, {Vreeswijk},
  {Khazov}, {Fox}, {Cao}, {Gnat}, {Kelly}, {Nugent}, {Filippenko}, {Laher},
  {Wozniak}, {Lee}, {Rebbapragada}, {Maguire}, {Sullivan}, \&
  {Soumagnac}}]{Yaron_2017}
{Yaron}, O., {Perley}, D.~A., {Gal-Yam}, A., {et~al.} 2017,
  \href{http://dx.doi.org/10.1038/nphys4025}{\JournalTitle{Nature Physics}, 13,
  510}

\bibitem[{{Yoon} \& {Cantiello}(2010)}]{Yoon2010}
{Yoon}, S.-C., \& {Cantiello}, M. 2010,
  \href{http://dx.doi.org/10.1088/2041-8205/717/1/L62}{\JournalTitle{\apjl},
  717, L62}

\end{thebibliography}

\end{document}